\documentclass[12pt,dvips]{article}

\usepackage{amsmath,amssymb,exscale}
\usepackage{array,multicol}
\usepackage{afterpage,float,flafter}
\usepackage{epsfig,rotating,pifont}
\usepackage{cite}
\title{
\vspace{1cm}
\Large\textbf{Supersymmetry breaking\\ by Wilson lines in AdS$_5$}
\vspace*{.5cm}
\author{\large \textbf{
Jonathan Bagger\footnote{email: bagger@jhu.edu}
\mbox{  }and Michele Redi\footnote{email: redi@pha.jhu.edu}}\\
\emph{
Department of Physics and Astronomy} \\
\emph{Johns Hopkins University} \\
\emph{3400 North Charles Street} \\
\emph{Baltimore, MD 21218-2686}}}

\date{}
\begin{document}
\maketitle
\thispagestyle{empty}
\vspace*{.5cm}

\begin{abstract}
In the Randall-Sundrum compactification of AdS$_5$ with detuned
brane tensions, supersymmetry can be spontaneously broken
by a non-trivial Wilson line for the graviphoton.
The supersymmetry breaking vanishes in the tuned limit.  This
effect is equivalent to supersymmetry breaking by Scherk-Schwarz
boundary conditions.
\end{abstract}

\newpage
\renewcommand{\thepage}{\arabic{page}}
\setcounter{page}{1}

In this note we show that in the supersymmetric Randall-Sundrum
model \cite{rs} with detuned brane tensions \cite{kaloper},
supersymmetry can be broken by the Hosotani (or Wilson line)
mechanism \cite{hosotani}. The setup is based on five-dimensional
anti de-Sitter (AdS$_5$) supergravity, compactified on the
orbifold $S^1/\mathbb{Z}_2$, with branes located at the orbifold
fixed points.  The brane actions and supersymmetry transformations
are adjusted so that the bulk-plus-brane theory is locally
supersymmetric.  This can always be done provided the magnitudes
of the brane tensions do not exceed the tuned value \cite{bagger}.

In five-dimensional AdS supergravity, the dynamical fields are the
vielbein $e_M^A$, the graviphoton $B_M$, and a symplectic Majorana
gravitino $\Psi_{Mi}$. When the tensions are tuned, the
ground-state metric on each brane is flat, and the distance
between the branes is a modulus of the compactification. In the
detuned case, the metric on each brane is AdS$_4$, and the
distance between the branes is fixed in terms of the brane
tensions and the bulk cosmological constant. In each case, the
vacuum expectation value (VEV) of $B_5$ is not determined by the
classical equations of motion.\footnote{Note that the VEV of $B_5$
is gauge invariant for gauge transformations defined on the
circle. A non-zero VEV of $B_5$ gives rise to a non-trivial Wilson
line for the graviphoton around the fifth dimension.} It is a
modulus of the compactification, even in the detuned case.

The graviphoton gauges a $U(1)$ subgroup of the flat-space $SU(2)$
automorphism group.  Under this $U(1)$, the gravitino is charged,
with charge proportional to the AdS$_5$ curvature. As in the
ordinary Hosotani mechanism, the VEV of $B_5$ induces a gravitino
bilinear in the five-dimensional action.\footnote{This mechanism
is reminiscent of one proposed in \cite{quiros}, where an
auxiliary field was used to trigger supersymmetry breaking in flat
five-dimensional supergravity.}  At first glance, one might think
that supersymmetry is spontaneously broken.  In fact, we will see
that a VEV breaks supersymmetry when the branes are not tuned, but
does not break supersymmetry when they are.  This is related to
the fact that the distance between the branes is a modulus in the
tuned limit.

In ref.\ \cite{bagger2}, using the interval representation of the
orbifold, boundary conditions consistent with local supersymmetry
were found that spontaneously break global supersymmetry. The
boundary conditions were proved to be equivalent to a
Scherk-Schwarz twist on the orbifold covering space. In the final
part of this paper, we will show that these results are equivalent
to the breaking by a Wilson line.

We start recalling the action for pure ${\cal N}=2$, $D=5$
supergravity with cosmological constant,
\begin{eqnarray}
S_{\rm bulk}&=&M_5^3\int d^5x~e_5 \Big[-\frac 1 2 R+6
k^2-\frac 1 4 F^{MN}F_{MN}+\frac i 2 \bar{\Psi}_{Mi} \Gamma^{MNK}D_N \Psi_{Ki}\nonumber \\
&&-\frac 3 2 k~\bar{\Psi}_{Mi} \Sigma^{MN} \vec q \cdot \vec
\sigma_{ij}\Psi_{Nj}+\frac {\sqrt{6}} 4 k~B_N
\bar{\Psi}_{Mi}\Gamma^{MNK}\vec q \cdot \vec
\sigma_{ij}\Psi_{Kj} \nonumber \\
&-&i\frac {\sqrt{6}} {16}
F_{MN}\Big(2\bar{\Psi}^M_i\Psi^N_i+\bar{\Psi}_{Pi}\Gamma^{MNPQ}\Psi_{Qi}\Big)-\frac
1 {6\sqrt{6}} \epsilon^{MNPQR}F_{MN}F_{PQ}B_R\Big],
 \label{bulk}
\end{eqnarray}
where $M_5$ is the five-dimensional Planck mass and $k$ is
the bulk cosmological constant.  The gravitino $\Psi_{Mi}$ is a
symplectic Majorana spinor and
$\vec q=(q_1,~q_2,~q_3)$ is a constant unit vector that parameterizes
the gauged $U(1)$ subgroup of $SU(2)$.  As the notation suggests,
$\Psi_{Mi}$ transforms as a doublet under $SU(2)$, but the $U(1)$
gauge coupling explicitly breaks $SU(2)$ down to $U(1)$.  Under
the $U(1)$ gauge symmetry, the fields transform as
\begin{eqnarray}
\Psi_{Mi} &\to& \exp\Big[i k \sqrt{\frac 3 2}~ \vec q \cdot \vec
\sigma \lambda\Big]_{ij}~\Psi_{Mj}
\nonumber \\[2mm]
B_M&\to& B_M+\partial_M \lambda.
 \label{gauge}
\end{eqnarray}
The gravitino charge is proportional to the five-dimensional
cosmological constant, $k$.

To compactify the theory on the orbifold $S^1/\mathbb{Z}_2$, we
decompose the symplectic Majorana spinor $\Psi_i$ in two-component
notation,
\begin{equation}
\Psi_{1}=\left(\begin{matrix} ~~\psi_{1\alpha} \\
~~\bar{\psi}_{2}^{\dot{\alpha}}
\end{matrix} \right),~~~~~~~~~~\Psi_{2}=\left(\begin{matrix} -\psi_{2\alpha} \\
~~\bar{\psi}_1^{\dot{\alpha}}
\end{matrix} \right).
\end{equation}
We also assign the following parities to the fields,

\begin{center}
\begin{tabular}{cccccccc}
even:&&$\quad e_m^a$, &$\quad e_5^{\hat{5}}$, &$\quad B_5$, &$\quad \psi_{m1}$, &$\quad \psi_{52}$ \\
odd:&&$\quad e_m^{\hat{5}}$,&$\quad B_m$, &$\quad \psi_{m2}$,&$\quad \psi_{51}$.
\end{tabular}
\end{center}

\noindent
With these choices, the parameter $q_3$ must have
odd parity for the action to be $\mathbb{Z}_2$
invariant.

We work in the orbifold covering space, with the two branes
of tensions $T_0$ and $T_\pi$, located at
$\phi=0$ and $\phi=\pi$ along the fifth-dimension.  Local
supersymmetry requires that we include the following
brane action \cite{bagger}
\begin{eqnarray}
S_{brane}=&-&\int d^4xd\phi~
e_4[T_0+2\alpha_0(\psi_{m1}\sigma^{mn}\psi_{n1}+h.c)]\delta(\phi)\nonumber\\
&-&\int d^4xd\phi~
e_4[T_\pi-2\alpha_\pi(\psi_{m1}\sigma^{mn}\psi_{n1}+h.c)]\delta(\phi-\pi),
\label{branes}
\end{eqnarray}
where the brane-localized gravitino masses are fixed by
the brane tensions and the vector $\vec q$. (For related
work, see \cite{old}).

The ground state metric is
\begin{equation}
ds^2=F(\phi)^2g_{mn}dx^m dx^n +r_0^2 d\phi^2,
\label{vacuum}
\end{equation}
where the warp factor is
\begin{equation}
F(\phi)=e^{-k r_0 |\phi|}+\Big(\frac {T-T_0}{T+T_0}\Big) e^{k r_0
|\phi|}
\end{equation}
and $T$ is the tuned tension,
\begin{equation}
T=6M_5^3 k.
\end{equation}
The metric $g_{mn}$ in eq.\ (\ref{vacuum}) is AdS$_4$ with radius
of curvature $L$, where
\begin{equation}
\frac 1 {4 k^2 L^2}=\frac {T-T_0} {T+T_0}.
\end{equation}
The radius of the orbifold is fixed to be
\begin{equation}
r_0=\frac 1 {2\pi k}\log\frac
{(T+T_0)(T+T_\pi)}{(T-T_0)(T-T_\pi)}. \label{radius}
\end{equation}
Note that when only one tension is tuned, the critical distance
between the branes is infinite, while if the tensions are equal
and opposite, the distance is zero.  In the tuned case, with
$T_0=-T_\pi=T$, the radius is not determined.

The full bulk-plus-brane action is invariant under
five-dimensional ${\cal N}=2$ supersymmetry, restricted to ${\cal
N}=1$ at the orbifold fixed points.  In four-dimensional language,
the variations of the fermionic fields are \cite{bagger},
\begin{eqnarray}
\delta \psi_{m1}&=& 2 D_m \eta_1+i k\sigma_m (q_3
\bar{\eta}_2+q_{12}^*\bar{\eta}_1)-\frac 2 {\sqrt{6}}i e_5^{\hat
5}
F^{n5}(\sigma_{mn}+g_{mn})\eta_1 \nonumber \\
\delta \psi_{m2}&=& 2 D_m \eta_2+i k \sigma_m (q_3
\bar{\eta}_1-q_{12}\bar{\eta}_2)-\frac 2 {\sqrt{6}} i e_5^{\hat 5}
F^{n5}(\sigma_{mn}+g_{mn})\eta_2 \nonumber \\
\delta \psi_{51}&=& 2 D_5\eta_1+k(e_5^{\hat{5}}-i
\sqrt{6}B_5)(q_3\eta_1-q_{12}^* \eta_2)-\frac 2
{\sqrt{6}}F_{m5}\sigma^m \bar{\eta}_2
\nonumber \\
\delta \psi_{52}&=&2 D_5\eta_2-k(e_5^{\hat{5}}-i
\sqrt{6}B_5)(q_3\eta_2+q_{12}\eta_1)+\frac 2
{\sqrt{6}}F_{m5}\sigma^m \bar{\eta}_1-4(\alpha_0
\delta(\phi)-\alpha_\pi \delta(\phi-\pi))\eta_1,\nonumber \\
\label{variations}
\end{eqnarray}
where $q_{12}=q_1+i q_2$, and for simplicity, we have set
$e_m^{\hat{5}}=B_m=0$.  The supersymmetry parameters $\eta_{1,2}$
are even and odd, respectively.  Note that the supersymmetry
transformation $\delta \psi_{52}$ contains a brane-localized
contribution.

To discuss supersymmetry breaking, we consider the Killing spinor
equations following from (\ref{variations}).  The equations can
be solved by separation of variables, taking
$\eta_{1,2}(x,\phi)=\beta_{1,2}(\phi)\eta(x)$.  The conditions
$\delta\psi_{m1}=0$ and $\delta \psi_{m2}=0$ imply that the
four-dimensional spinor $\eta(x)$ satisfies the four-dimensional
Killing spinor equation in AdS$_4$. The equations $\delta
\psi_{51}=0$ and $\delta \psi_{52}=0$ constrain the fermionic warp
factors $\beta_{1,2}(\phi)$. In a constant $B_5$ background, we
find
\begin{eqnarray}
2\partial_5\beta_1+k(r_0-i\sqrt{6}B_5)(q_3 \beta_1-q_{12}^*
\beta_2)&=&0
\nonumber \\
2\partial_5\beta_2-k(r_0-i\sqrt{6}B_5)(q_3 \beta_2+q_{12}
\beta_1)&=&4(\alpha_0 \delta(\phi)-\alpha_\pi
\delta(\phi-\pi))\beta_1. \label{fermionicwarp}
\end{eqnarray}

To simplify the discussion we restrict our attention to the case
$q_3=\epsilon(\phi)$, $q_{1,2}=0$, where $\epsilon(\phi)$ is the
step function.  (Any other choice is physically equivalent and can
be obtained by an $SU(2)$ rotation of the gravitino.) The
solution to the above equations, defined on the orbifold, is
\begin{eqnarray}
\beta_1(\phi)&=& A~\exp\Big[-\frac k 2(r_0-i \sqrt{6} B_5)|\phi|\Big]\nonumber \\
\beta_2(\phi)&=&B~\exp\Big[+\frac k 2(r_0-i \sqrt{6}
B_5)|\phi|\Big]\epsilon(\phi).
\end{eqnarray}
In general, the $\beta_{1,2}$ are solutions {\it except} at the
orbifold fixed points. Global supersymmetry demands that the
Killing spinors be solutions everywhere. This requires that we
match the delta-function singularities, which gives
\begin{eqnarray}
\frac B A&=&\alpha_0 \nonumber \\[3mm]
\frac B A&=&\alpha_\pi \exp\big[-\pi k(r_0-i\sqrt{6}B_5)\big].
\label{constraints}
\end{eqnarray}
Local supersymmetry determines the magnitudes of coefficients
($\alpha_0$,\ $\alpha_\pi$) in terms of the brane tensions
\cite{bagger},
\begin{eqnarray}
|\alpha_0|^2&=&\frac {T-T_0}{T+T_0}\nonumber\\[3mm]
|\alpha_\pi|&=&|\alpha_0| \exp(k \pi r_0).
\end{eqnarray}

When $\alpha_0$ and $\alpha_\pi$ have the same phase, it follows
from (\ref{constraints}) that unbroken supersymmetry requires
$B_5=0$. Then, for generic values of $B_5$, eqs.\
(\ref{constraints}) have no solution, and supersymmetry is
spontaneously broken.  When $\alpha_0$ and $\alpha_\pi$ have
different phases, the situation is the same, up to a shift in the
origin of $B_5$. Note that the second of eqs.\ (\ref{constraints})
shows that $B_5$ is a periodic variable,
\begin{equation}
\big[B_5\big]=\frac {2} {\sqrt{6}k}.
\end{equation}
The supersymmetry breaking is invariant under shifts by
$2/(\sqrt{6}k)$.

The tuned model is a degenerate case.  Since $\alpha_{0,\pi}=0$,
eqs.\ (\ref{constraints}) are always satisfied, and the VEV of
$B_5$ does not break supersymmetry.  A simple argument explains
why this must be so.  When the tensions are tuned, the radius
of the orbifold is a modulus.  In the low-energy effective theory,
as shown in \cite{radion}, the superpotential for the radion
vanishes, which implies that supersymmetry cannot be broken.  In
contrast, in the detuned case, the superpotential for the radion
is non-zero \cite{us}.  Even though $B_5$ is still a flat direction
of the potential, its VEV breaks supersymmetry and contributes to
the mass of the four-dimensional gravitino.

Supersymmetry breaking by the Hosotani mechanism can be
easily translated into the language of Scherk-Schwarz
compactification \cite{ss}.  From (\ref{gauge}), we see that a VEV
for $B_5$ can be gauged away by a non-periodic gauge
transformation, with parameter
\begin{equation}
\lambda=-\int_{0}^\phi B_5~d\phi^\prime .
\end{equation}
The gravitino transforms into
\begin{equation}
\Psi_{M1}\to \Psi_{M1}^\prime=W(\phi)\Psi_{M1},
\label{gravitino}
\end{equation}
where we have introduced the Wilson line $W(\phi)$,
\begin{equation}
W(\phi)=\exp\Big[-i k \sqrt{\frac 3 2}\int_{0}^\phi B_5
~d\phi^\prime\Big]. \label{wilson}
\end{equation}
For a non-trivial Wilson line, the gravitino becomes multivalued
on the circle.  In this picture,\break supersymmetry is broken by
non-periodic boundary conditions \cite{bagger2}.  Note that when
$B_5$ is quantized in units of $2/(\sqrt{6}k)$, the transformation
(\ref{gravitino}) does not change the periodicity of $\Psi_M$, so
the Wilson line can be absorbed by a redefinition of the gravitino
field.

In conclusion, we would like to comment on supersymmetry breaking
when vector- and hyper-multiplets are minimally coupled to
supergravity. The vector multiplet contains a gauge field $V_M$,
two symplectic Majorana spinors $\lambda_i$, and a real scalar.
The hypermultiplet consists of a Dirac spinor and two complex
scalars $H_i$.  The fields $\lambda_i$ and $H_i$ transform as
doublets under $SU(2)$; all other fields are singlets. In AdS
supergravity, since a $U(1)$ subgroup of $SU(2)$ is gauged, the
graviphoton couples to $\lambda_i$ and $H_i$. As a result, the
Wilson line supersymmetry breaking is communicated to this sector.
This effect persists even when gravity is decoupled by taking
$M_5\to\infty$ with $L$ is finite.

\section*{Acknowledgments}
We thank D. Belyaev and A. Delgado for useful discussions. This
work was supported in part by the U.S. National Science
Foundation, grant NSF-PHY-9970781.

\end{document}